\documentclass[twocolumn]{aastex631}
\usepackage{amsmath}
\usepackage{hyperref}
\usepackage{soul}
\usepackage{graphicx} 

\received{2023 June 12}
\revised{2023 September 1}
\accepted{2023 September 3}
\submitjournal{ApJ Letters}

\begin{document}

\title{Constraints on the intergalactic magnetic field strength from $\gamma$-ray observations of GRB 221009A}

\author[0000-0002-6036-985X]{Yi-Yun Huang}
\affiliation{School of Astronomy and Space Science, Nanjing University, Nanjing 210093, China; xywang@nju.edu.cn; hmzhang@nju.edu.cn; ryliu@nju.edu.cn}
\affiliation{Key Laboratory of Modern Astronomy and Astrophysics (Nanjing University), Ministry of Education, Nanjing
210093, China}

\author[0000-0002-0170-0741]{Cui-Yuan Dai}
\affiliation{School of Astronomy and Space Science, Nanjing University, Nanjing 210093, China; xywang@nju.edu.cn; hmzhang@nju.edu.cn; ryliu@nju.edu.cn}
\affiliation{Key Laboratory of Modern Astronomy and Astrophysics (Nanjing University), Ministry of Education, Nanjing
210093, China}

\author[0000-0001-6863-5369]{Hai-Ming Zhang}
\affiliation{School of Astronomy and Space Science, Nanjing University, Nanjing 210093, China; xywang@nju.edu.cn; hmzhang@nju.edu.cn; ryliu@nju.edu.cn}
\affiliation{Key Laboratory of Modern Astronomy and Astrophysics (Nanjing University), Ministry of Education, Nanjing
210093, China}

\author[0000-0003-1576-0961]{Ruo-Yu Liu}
\affiliation{School of Astronomy and Space Science, Nanjing University, Nanjing 210093, China; xywang@nju.edu.cn; hmzhang@nju.edu.cn; ryliu@nju.edu.cn}
\affiliation{Key Laboratory of Modern Astronomy and Astrophysics (Nanjing University), Ministry of Education, Nanjing
210093, China}
\author[0000-0002-5881-335X]{Xiang-Yu Wang}
\affiliation{School of Astronomy and Space Science, Nanjing University, Nanjing 210093, China; xywang@nju.edu.cn; hmzhang@nju.edu.cn; ryliu@nju.edu.cn}
\affiliation{Key Laboratory of Modern Astronomy and Astrophysics (Nanjing University), Ministry of Education, Nanjing
210093, China}

\begin{abstract}

Characteristics of the cascade gamma-ray signal resulting from very-high-energy gamma-ray sources, such as gamma-ray bursts, can be used to constrain the strength and structure of intergalactic magnetic fields (IGMF). There has been a debate on whether GRB 190114C, the first gamma-ray burst with observed TeV photons, can constrain the IGMF. Recently, LHAASO detected the brightest-of-all-time GRB 221009A, which has much larger energy in TeV band and the spectrum extends to energy above 10 TeV, providing an unprecedented opportunity to studying IGMF. We perform a Monte-Carlo simulation of the cascade process with the public ELMAG code, considering the  TeV data of GRB 221009A observed by LHAASO. By comparing the resulting cascade emission with the flux limit obtained from Fermi-LAT observations, we infer  a limit of $B\ge 10^{-18.5}\rm G$ for IGMF. 
Though this limit may  not be as strong as the limit from blazars, it serves as an independent constraint on IGMF from a new class of  TeV sources.

\end{abstract}


\section{Introduction} 
\label{sec:intro}

The magnetic fields in galaxies and galaxy clusters are thought to result from the amplification of seed magnetic fields, which might exist in their initial form in the intergalactic medium\citep{neronov_2010}.  There are two broad classes of models for their origin of the seed magnetic fields\citep{durrer_2013}: (1) cosmological models, in which the seed fields are generated in the early universe before the structure formation; (2) astrophysical models, in which the seed fields are produced during the epochs accompanying the gravitational collapse leading to structure formation. Measurements of the strength of intergalactic magnetic fields (IGMF) can provide an important clue on the origin of the initial seed fields. 


Very high energy (VHE, $\ge$100 GeV) transient sources such as flaring active galactic nuclei (AGNs) and gamma-ray bursts (GRBs) are viable tools to constrain IGMF \citep{plaga_1995,dai_2002,wang_2004,razzaque_2004,ichiki_2008}. During propagating in the universe, TeV photons emitting from these sources will interact with Extragalactic Background Light (EBL) and produce $e^+e^-$ pairs via pair production process. The created pairs are deflected by the IGMF and radiate secondary GeV-TeV emission through inverse Compton scattering (ICS) off Cosmic Microwave Background (CMB) photons. When these cascade photons arrive at Earth, their properties, such as the spectrum and the time delay with respect to the primary emission, carry critical information of the IGMF, and can be thus used as a diagnosis of the IGMF   \citep{dai_2002,wang_2004,Dermer_2011, taylor_2011}.

This method was first applied to blazars \citep{neronov_2010}, which give a lower bound on the IGMF at the lever of B$\ge 10^{-16}$ G from the non-detection of  GeV gamma-ray emission from electromagnetic cascade initiated by the primary TeV gamma-rays in intergalactic medium. 
However, the constraints were subsequently found to be subject to significant systematic effects, such as the unknown duty cycle \citep{Dermer_2011,taylor_2011}, the
poorly constrained spectral properties of the source, and uncertainties in the EBL spectrum \citep{Arlen_2014,Finke_2015}. 
For example, restricting TeV activity of 1ES 0229+200 to a timescale of 3-4 years during which the source has been observed leads to a more robust lower limit of  B$\ge 10^{-18}$ G \citep{Dermer_2011}. Under such circumstances,  GRBs, as an independent class of transient TeV sources, becomes crucial for constraining IGMF.

As a short-lived powerful TeV source, GRB is suitable for studying IGMF because the time-delayed cascade photons can be easily distinguished from the primary photons. GRB 190114C is the first GRB observed with TeV emission and  a limit of $\rm B\ge 10^{-19.5}G$ on IGMF (for the coherence length of $ \lambda \le$ 1 Mpc) was inferred \citep{wang_2020}. On the other hand, \cite{dzhatdoev_2020}  used Monte-Carlo code ELMAG to calculate the cascade emission for various EBL models, and found that the sensitivity of Fermi/LAT is not sufficient to constrain the IGMF. The discrepancy could be due to that \cite{dzhatdoev_2020} only take into account
the primary TeV photons during the period from 62 s to
2454 s, while \cite{wang_2020} consider the power-law decay of the
afterglow flux starting from 6 s \citep{MAGIC_2019,Ravasio_2019,Wang_2019}, which  leads to a difference in the fluence of the primary TeV photons by a factor of 5.
Recently, LHAASO detected the brightest-of-all-time GRB 221009A\citep{LHAASO_2023}, which has much higher fluence in TeV band and the spectrum extends to energy above 10 TeV\citep{2022GCN.32677....1H}, thus it offers a precious opportunity for us to constrain IGMF.

LHAASO detected more than 60,000 photons with energies greater than $\sim 200\,{\rm GeV}$ from this GRB\citep{LHAASO_2023}. The observed spectra at various time intervals show sharp steepening at high energies due to the EBL absorption. The intrinsic spectra after correcting for EBL absorption can be described by a single power-law extending to the highest observed energy. By integrating the time-resolved intrinsic spectra, the isotropic equivalent energy in TeV is $E_{(\rm 0.3-15TeV)} \approx 1\times10^{53}$ erg. In addition, the clear steepening of the flux observed by LHAASO in the decay phase, which is consistent with a jet break, provides us the information regarding the half-opening angle ($\theta_j= 0.8\degr$). 

The rest of the paper is organized as follows. In \S 2, we describe our analysis of Fermi-LAT data in order to set experimental upper limits on the pair echo intensity from GRB 221009A. 
In \S 3, we use the open code ELMAG to study the electromagnetic cascades of primary TeV photons  and obtain constraints on IGMF. 
Finally we give conclusions in \S 4.

\section{Fermi-LAT data analysis}
At 13:16:59.99 UT ($\rm T_0$) on 2022 October 9, the Fermi Gamma-ray Burst Monitor (GBM) triggered and located GRB 221009A \citep{2022GCN.32636....1V,GBM_2023}, which was also detected by the Fermi Large Area Telescope (Fermi/LAT; \citealt{2022GCN.32637....1B}), Konus-Wind \citep{2022GCN.32668....1F}, Swift-BAT-GUANO \citep{2022GCN.32632....1D}, GECAM-C\citep{2022GCN.32751....1L,Yang2023,2023arXiv230301203A} and LHAASO\citep{LHAASO_2023}. Its  redshift is reported to be z = 0.151 \citep{2022GCN.32648....1D}.

The Fermi-LAT extended type data for the GRB 221009A are taken from the Fermi Science Support Center\footnote{\url{https://fermi.gsfc.nasa.gov}}. 
Only the data within  $30\degr \times 30\degr$ region of interest (ROI) centered on the position ($\rm R.A.=288.21\degr$,$\rm Decl.=19.73\degr$) of GRB 221009A are considered for the analysis.

We perform a binned maximum likelihood analysis for this GRB, and considering the LAT $SOURCE$ events between 100 MeV and 1 TeV. The corresponding instrument response function (IRF) (\textit{$P8R3\_SOURCE\_V$}3)\footnote{\url{https://fermi.gsfc.nasa.gov/ssc/data/analysis/documentation/Cicerone/Cicerone_Data/LAT_DP.html}} is used.
A maximum zenith angle of 90$\degr$ is adopted to reduce the contamination from the $\gamma$-ray Earth limb.
The LAT 12-year Source Catalog (4FGL-DR3) sources are included in our analysis.
For the main background component, we consider the isotropic emission template (``$iso\_P8R3\_SOURCE\_V3\_v1.txt$'') and the diffuse Galactic interstellar emission template (IEM; $gll\_iem\_v07.fits$) in our analysis.
The parameters of isotropic emission and IEM are left free.

Assuming a power-law spectrum ($dN/dE = AE^{\Gamma}$) for this burst, we obtained its spectral energy distribution (SEDs) of different time intervals with software $fermipy$ (version v1.1) \citep{2017ICRC...35..824W}. 
For the energy bins with a test statistic (TS\footnote{TS$= 2 (\ln\mathcal{L}_{1}-\ln\mathcal{L}_{0})$, where $\mathcal{L}_{1}$ and $\mathcal{L}_{0}$ are maximum likelihood values for the background with the GRB and without the GRB (null hypothesis).}) value $<$ 9 (TS=9 corresponding to $3\sigma$ significance), we extract the upper limits at 95\% confidence level. The SEDs of different time intervals are shown as the black points in Figures \ref{fig:MC_sed}, \ref{fig:400GeV}and \ref{fig:analytic}.

\section{Monte Carlo Simulation approach} \label{sec:Simulation}

\subsection{Monte Carlo Simulation}
\begin{figure}
    \centering
    \includegraphics[angle=0,scale=0.35]{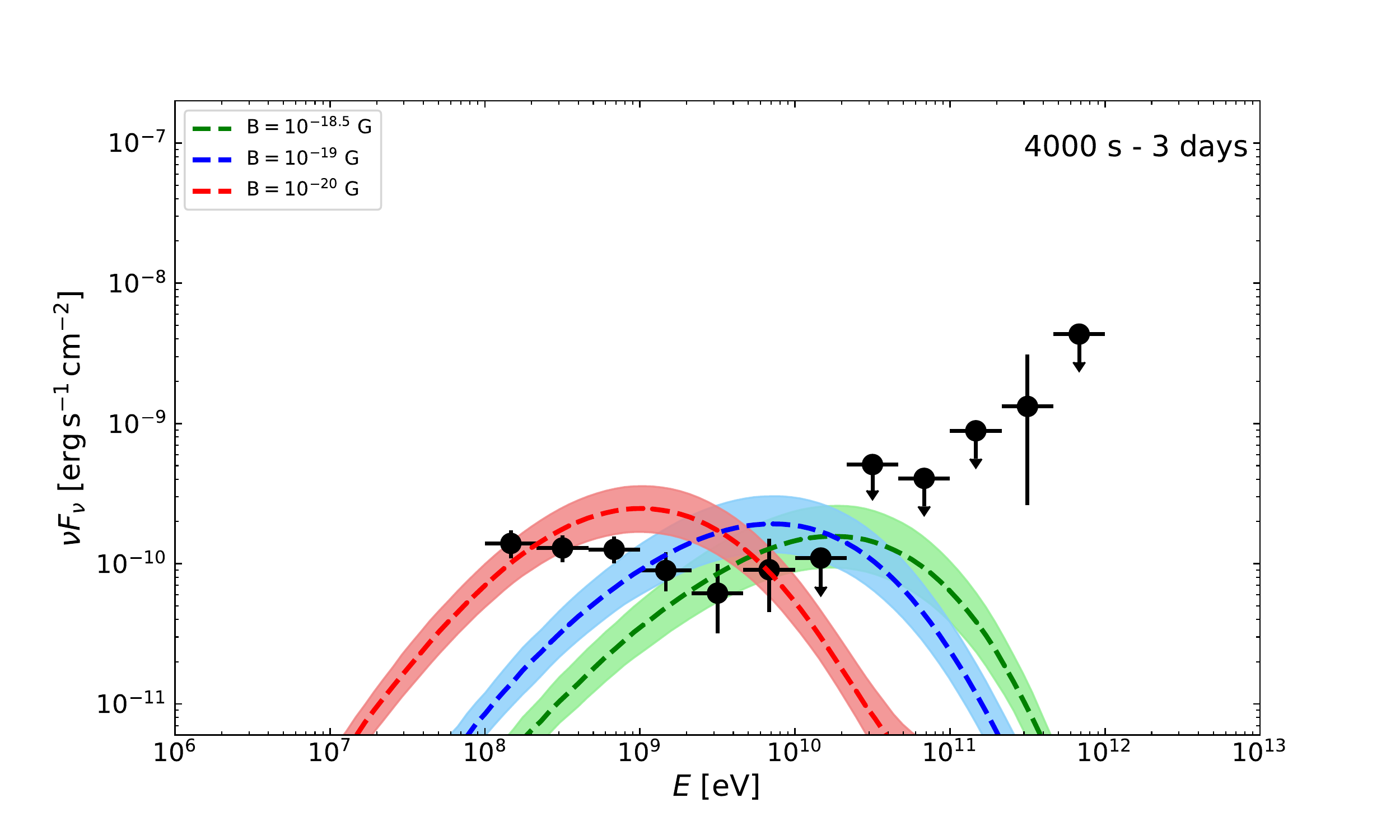}
    \includegraphics[angle=0,scale=0.35]{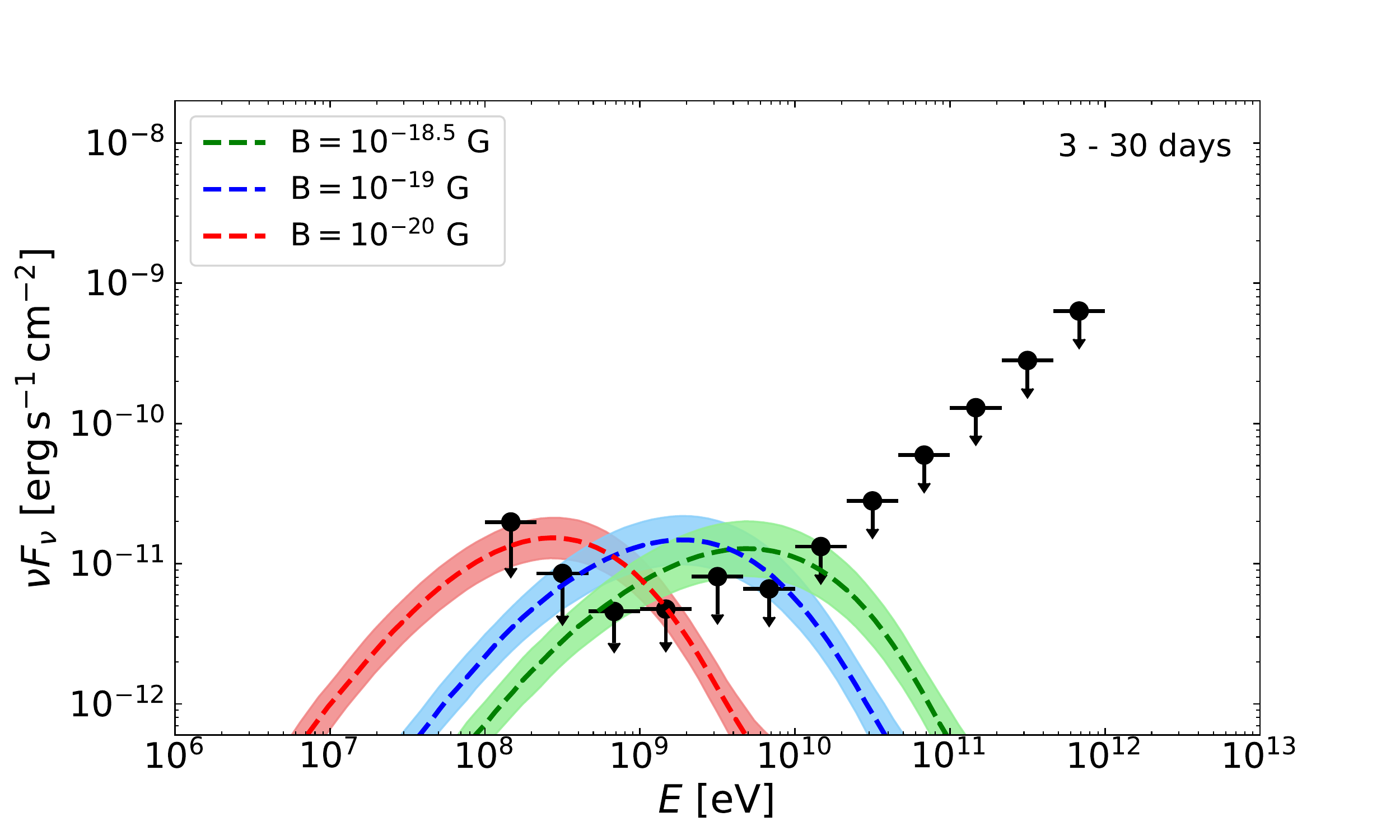}
    \includegraphics[angle=0,scale=0.35]{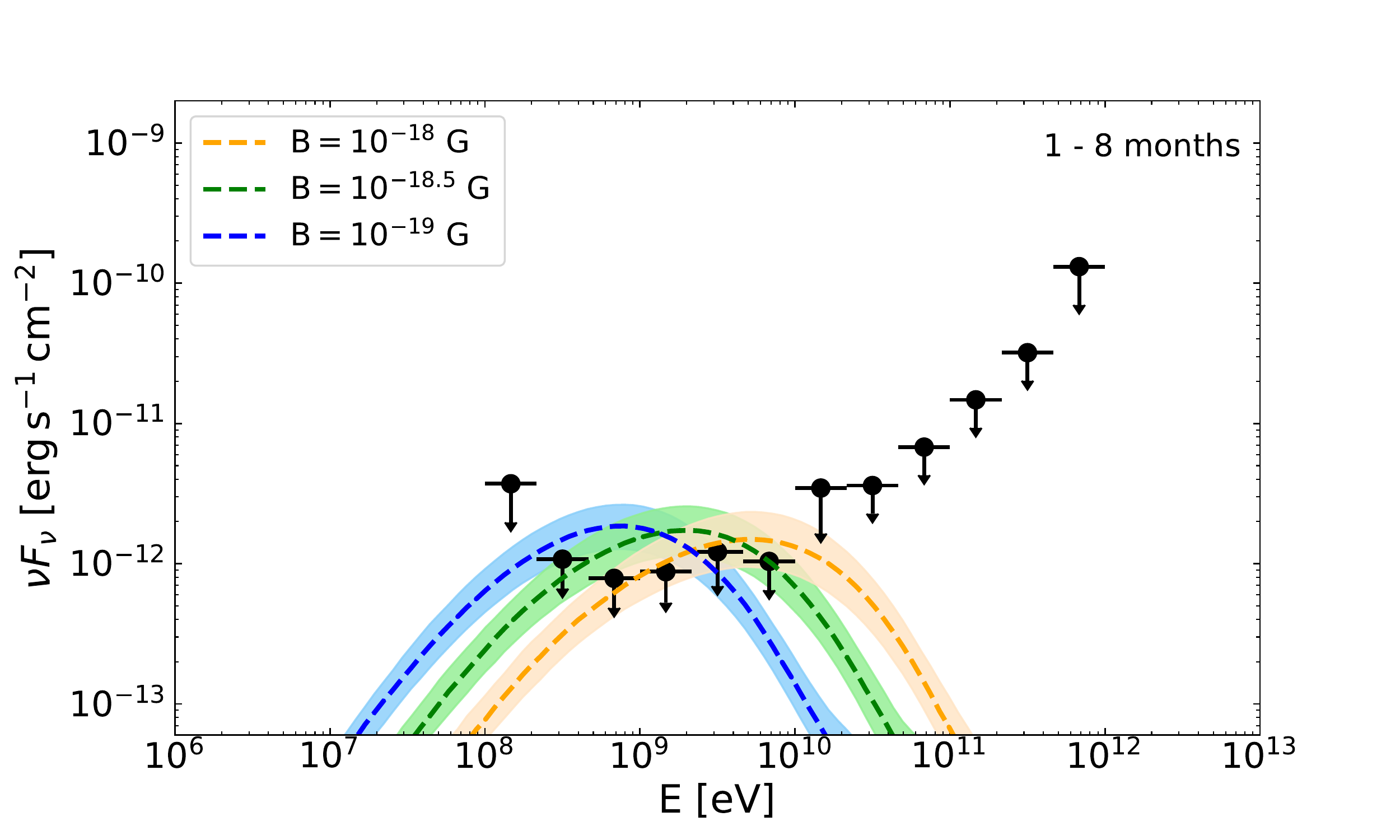}
    \caption{Comparison between the expected flux of the echo emission and the observed data by  Fermi/LAT in Monte-Carlo simulation. The dotted lines represent the echo emission using the standard flux of the EBL model of \citet{saldana_2021}, while the color bands represent the echo emission considering the uncertainties in the EBL intensity.}
    \label{fig:MC_sed}
\end{figure}
Very high energy photons are absorbed by EBL when they are propagating in the IGM, creating electrons and positrons through pair production. These pairs then scatter CMB photons to the GeV domain via inverse Compton (IC) radiation. Concurrently, the IGMF deflects the pairs, causing these secondary GeV photons to reach the observer with different directions and arrive later compared to the TeV photons. The characteristics of these secondary GeV photons, in terms of their duration and strength, can be thus used to constrain the IGMF. It is worth noting that the angular spread of pair production and IC emission induces an intrinsic time delay even  in the absence of the intervening magnetic fields. As estimated by \cite{Vovk2023} for GRB 190114C, this time spread can span a range  from $10^3$ to $10^5$ seconds depending on the energy. However, in this study, the effect of the intrinsic time spread is negligible due to the much longer time intervals used in this study.

The open code ELMAG \citep{kachelries_2012} is a Monte Carlo simulation program designed to study electromagnetic cascades on the EBL, including the deflections of charged particles in IGMF. Due to the introduction of some additional features, this program provides an accurate description of the particle trajectories. These features include the turbulence of extragalactic magnetic fields, the opening angle of jet and the calculation of three-dimensional trajectories of the secondary electrons and positrons  by solving the Lorentz force equation. In this work, we use the newest edition ELMAG 3.03 \citep{blytt_2020} to perform a full three-dimensional simulation.

We choose three time intervals: $T_0$+4000 seconds to 3 days, $T_0$+3 days to 30 days and $T_0$+30 days to 8 months to probe an early stage and two longer observation windows. Here, the 4000 s is the moment when the Fermi-LAT detector entered the field of view for the second cycle. Furthermore, the 3rd day represents the transition time from Fermi-LAT detection to non-detection of GeV photons \citep{2023arXiv230303855S}. Moreover, to obtain a more stringent constraint on the IGMF, we set the observation limit at 30 days, considering that the theoretical cascade flux diminishes with time (i.e., decreases as $t^{-1}$) while the Fermi-LAT upper limit flux follows $t^{-1/2}$ for long-term observations. The final time bin extending to 8 months corresponds to the data accumulated as of writing of this paper. 

We use a turbulent magnetic field with a Kolmogorov spectrum{\footnote{Note that the bug in generation of the helical field in the Monte-Carlo code pointed out by \citet{Kalashev_2022}  has been corrected in ELMAG 3.03.}}, where the minimum and maximum spatial scales are set as the default values $\mathrm{L}_\mathrm{min} = 5\times 10^{-4} $ Mpc and $\mathrm{L}_\mathrm{max}$ = 5 Mpc, respectively.  A coherence length of 1 Mpc is assumed,  consistent with the analytic calculation. 

\begin{table}[ht!]
\caption{Input parameters used in the Monte-Carlo simulation for three EBL model. The intrinsic spectrum is assuming as a power-law function (i.e., $dN/dE = A(E/\mathrm{TeV})^{-\Gamma}$). }
\hspace{-1cm} 
\scalebox{0.7}{
    \begin{tabular}{lccc}
        \hline\hline 
        Parameters &\multicolumn{3}{c}{EBL models}\\
                   & \small Low EBL & \small Standard EBL & \small High EBL\\ 
        \hline\hline
        Normalization factor A ($\mathrm{TeV}^{-1} ~\mathrm{cm}^{-2} ~\mathrm{s}^{-1}$) &1.20$\times10^{-7}$ &1.56$\times10^{-7}$  &2.02$\times10^{-7}$\\
        Spectral index &2.5 &2.36  &2.22\\
        Minimal injection energy &\multicolumn{3}{c}{0.3 TeV}\\
        Maximal injection energy &\multicolumn{3}{c}{15 TeV}\\
        Redshift &\multicolumn{3}{c}{0.15}\\
        Magnetic field strength &\multicolumn{3}{c}{$\rm [10^{-20} G;10^{-19}G;10^{-18.5}G]$ }\\
        Coherence length &\multicolumn{3}{c}{1 Mpc}\\
        IGMF minimal spatial scale &\multicolumn{3}{c}{$5\times 10^{-4}$ Mpc}\\
        IGMF maximal spatial scale &\multicolumn{3}{c}{5 Mpc}\\
        Jet opening angle &\multicolumn{3}{c}{$1\degr$}\\
        Jet misalignment angle &\multicolumn{3}{c}{$0\degr$}\\
        \hline\hline
    \end{tabular}
}
\label{tab:1}
\end{table}{}

Due to the potential systematic effect existing in different EBL models, we follow the approach described in \cite{LHAASO_2023} to take three EBL models into consideration, namely, the models of weak attenuation, standard situation and strong attenuation in the \cite{saldana_2021} model. The corresponding intrinsic spectra, after the correction for these EBL models, have been obtained in \cite{LHAASO_2023}. To simplify the calculation, the spectral index is adopted from the value during the time interval when the TeV flux is {the highest}\footnote{The effect of the the uncertainties in the TeV flux normalization and slope and their time variation on the cascade flux is found to be unimportant as it is  smaller than that caused by the uncertainty in EBL models.}.
Then we perform the simulations for these three EBL models using the input parameters shown in Table.\ref{tab:1}. In our calculation, we treat the primary emission as an instantaneous injection to simplify the calculation. But for the first time interval, considering that the duration of primary TeV emission is comparable to this time interval, such simplification is less accurate. Therefore we calculate the start time for each time-resolved spectrum of the primary emission \citep{LHAASO_2023} and integrate their cascade emission over the time interval. The results are displayed in Figure \ref{fig:MC_sed}. Notably, the lower and upper ranges of the color bands in Figure \ref{fig:MC_sed} represent the results obtained from the low and high EBL intensity models, respectively. For IGMF ranging from $B=10^{-20}$ G to $B=10^{-18.5}$ G, the theoretical flux  significantly exceeds the Fermi-LAT upper limits in some energy ranges, while for the case of $B=10^{-18}$ G (as shown in the bottom panel of Figure \ref{fig:MC_sed}), the theoretical flux is below the upper limits  in the low EBL model case. We thus conclude that a conservative lower limit of the IGMF is $B\ge 10^{-18.5}$ G.  This limit on IGMF is  much more stringent  than that derived from GRB 190114C \citep{wang_2020,dzhatdoev_2020}.

\subsection{Constraining IGMF using analytic approach}
For a cross-check, we perform an analytic calculation of the cascade emission and then derive the constraint on IGMF. A comprehensive description  of the analytic approach is provided in the Appendix. We find that the constraint on IGMF through the analytic process is $\ge 10^{-18}$ Gauss. The analytic approach  does not take into account of the uncertainties in EBL models. Additionally, we explore the case where the maximum energy is 7 TeV, which is the maximum energy reported by WCDA observations. We find that the  constraint on IGMF is almost the same, as shown in the  appendix Figure \ref{fig:analytic}. 

\subsection{On the origin of the 398 GeV photon}
\begin{figure}
    \centering
    \includegraphics[angle=0,scale=0.35]{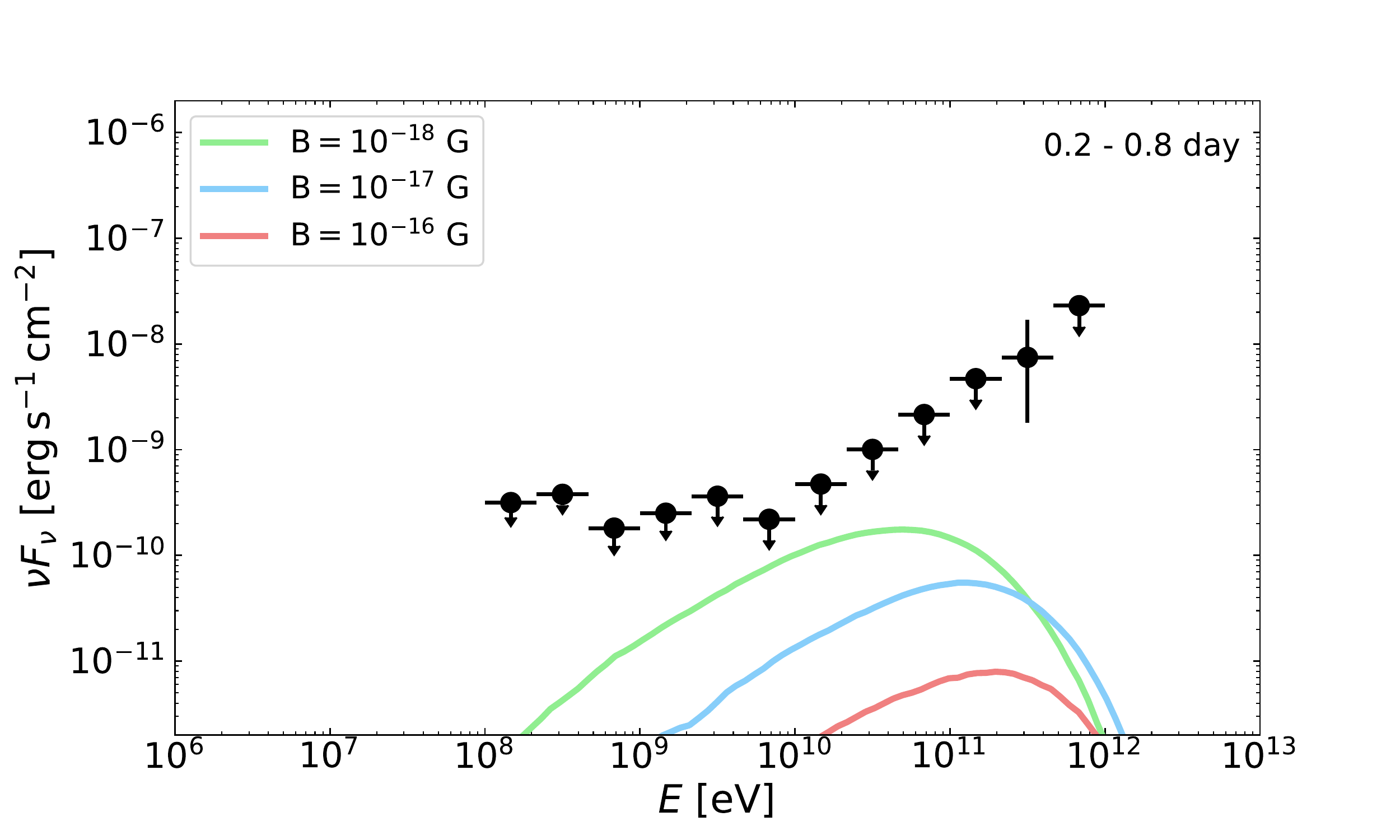}
    \caption{Comparison between the expected flux of the echo emission and the observed data by  Fermi/LAT in the time interval of  $\rm T_0+0.2$-$\rm T_0+0.8$ day, during which the 397 GeV photon was detected.}
    \label{fig:400GeV}
\end{figure}

\citet{xia_2022} proposed that a 398 GeV photon arriving at 0.4 day could come from the cascade emission. Under this hypothesis, it requires the strength of IGMF to be $\sim 10^{-17}\rm G$. In order to verify this scenario, we adopt the precise parameters of the TeV emission  measured by LHAASO to study the possibility. We use ELMAG to calculate the cascade flux within the energy range of 200 to 800 GeV over a period of from 0.2 to 0.8 days after the burst for three magnetic fields ($10^{-18}$ G,$10^{-17}$G and $10^{-16}$ G). For simplicity, we only consider the standard case of the EBL model in \citet{saldana_2021}. Figure.\ref{fig:400GeV} presents a comparison between the theoretical flux of the cascade emission and the observed data by Fermi-LAT. As we can see, the theoretical flux is much lower than the observed one at the energy around 398 GeV. By considering the corresponding exposure time and effective area of LAT during this event, we estimate the expected photon number. The Poisson probabilities for detecting one such photon  are 0.65\%,
0.82\%, 0.17\% for the three magnetic field cases, respectively. Although  the probability for such an event is low, we cannot conclusively rule out the scenario at a confidence level of $\ga3\sigma$. Nevertheless, should this scenario hold, it would require a magnetic field of approximately $10^{-17}$ G, which does not conflict with our lower boundary on IGMF.

\section{Discussion and conclusion} 
\label{sec:discussion}

GRB 221009A, as a once-in-ten-thousand-year event \citep{2023ApJ...946L..31B}, has the largest amount of energy in the TeV band and its spectrum extends to at least 10\,TeV \citep{2022GCN.32677....1H}. In this work, we obtain the constraints on the IGMF based on the LHAASO and Fermi-LAT observations of GRB 221009A. 
We used a Monte-Carlo code ELMAG 3.03 to calculate the cascade emission.  The constraints on magnetic field from the analytic calculation  are  consistent  with that from the Monte-Carlo code. The Monte-Carlo approach takes into account the uncertainty in the EBL intensity and we  obtain a conservative lower limit, which is  ${\rm B}\ge 10^{-18.5}$ G for the coherence length of $\lambda \le$1 Mpc. This result is much more stringent than that derived from GRB 190114C \citep{wang_2020,dzhatdoev_2020}. While this limit may be not as strong as that from blazars (see the recent work \citet{2023A&A...670A.145A}, which obtained a limit of $\ge 10^{-17}$ G ),  it serves as an independent constraint on IGMF from a new class of  TeV sources.


One significant systematic effect originates from uncertainties in the EBL model. Following the treatment in \cite{LHAASO_2023}, we adopted two additional EBL intensity models to testify the systematic error they may bring, corresponding to the lower (weak attenuation) and upper (strong attenuation) boundary of the error range in the \cite{saldana_2021} model. Most of other EBL models are compatible with this error band. We would like to point out that our limits are derived from the lowest EBL intensity in the model of \cite{saldana_2021}, which represnts a conservative constraint on IGMF.

Another parameter affecting the constraints on the IGMF strength is the coherence length, $\lambda_\mathrm{coh}$. In this work, we set the coherence length as 1 Mpc (approximately equal to the cooling length of electrons $\lambda_\mathrm{IC}$ which upscatter CMB photons to 1 GeV). As indicated by the equation of $\theta_{\rm B}(\lambda_\mathrm{coh}\le\lambda_\mathrm{IC})$ in the Appendix, the constraint will become more stringent as $\lambda_\mathrm{coh}$ decreases, scaling as $\lambda_\mathrm{coh}^{-1/2}$\citep{neronov_2010}, while the IGMF constraint will be independent of coherence length if $\lambda_\mathrm{coh}>{\rm 1 Mpc}$. 

\begin{acknowledgments}

We thank the anonymous referee for valuable suggestions, and we thank Ievgen Vovk for helpful discussions. The work is supported by the  National Key R$\&$D Program of China under grant No. 2022YFF0711404, the NSFC Grants No.12121003, No. 12203022 and No. U2031105, and China Manned Spaced Project (CMS-CSST-2021-B11).

\end{acknowledgments}

\bibliography{cascade}{}
\bibliographystyle{aasjournal}

\appendix
\section{Analytic approach} 
\label{sec:analytic}
\subsection{Description of the analytic approach} 
Here, we present an estimate of the cascade emission using an analytic approach, which is  useful for the understanding of the underlying physics.
The primary photons with energy range $>0.2$ TeV from GRB 221009A produce electron-positron pairs by photon-photon interactions with the EBL photons when they propagate through intergalactic medium (IGM), i.e $\gamma+\gamma_0\rightarrow{e^{+}+ e^{-}}$. Here $\gamma$ and $\gamma_0$ represent VHE photons and low-energy EBL photons, respectively. The probability of pairs produced in collisions of a $\gamma$-ray with dimensionless energy $\epsilon = \mathrm{h} \nu /\mathrm{m}_{\mathrm{e}} \mathrm{c}^ 2$ follows a more accurate distribution, which is  used in analytic calculation below and Monte-Carlo simulation in Section \ref{sec:Simulation},  given by \citep{zdziarski_1988}
\begin{align}
P\left(\gamma_{\mathrm{e}}, \epsilon\right)= &\int_{\epsilon / x_{\gamma}}^{\infty} \mathrm{d} \epsilon_{0} n_{0}\left(\epsilon_{0}\right) \frac{3 \sigma_{\mathrm{T}} \mathrm{c}}{4 \epsilon^{2} \epsilon_{0}}\left[r-(2+r) \frac{\epsilon}{\epsilon_{0} x_{\gamma}}\right. \notag \\ &
\left.+2\left(\frac{\epsilon}{\epsilon_{0} x_{\gamma}}\right)^{2}+2 \frac{\epsilon}{\epsilon_{0} x_{\gamma}} \ln \frac{\epsilon_{0} x_{\gamma}}{\epsilon}\right]
\end{align}
where  $\mathrm{x}_{\gamma}=4 \gamma_{\mathrm{e}} \gamma_{\mathrm{e}}^{\prime}, r=\left(\gamma_{\mathrm{e}} / \gamma_{\mathrm{e}}^{\prime}+\gamma_{\mathrm{e}}^{\prime} / \gamma_{\mathrm{e}}\right) / 2, \gamma_{\mathrm{e}}^{\prime}=\epsilon-\gamma_{\mathrm{e}}$, $\sigma_{\mathrm{T}}$ is the Thomson cross section, and $n_{0}\left(\epsilon_{0}\right)$ is the EBL number density at energy $\epsilon_{0}$. 
The electron energy spectrum created by pair production can further be written as 
\begin{equation}
\frac{\mathrm{d} N_{\mathrm{e},0}(\epsilon)}{\mathrm{d} \gamma_{\mathrm{e}}}=n_{\mathrm{Tev}} p\left(\gamma_{\mathrm{e}}, \epsilon\right),
\end{equation}
where $p\left(\gamma_{\mathrm{e}}, \epsilon\right)=2 P\left(\gamma_{\mathrm{e}}, \epsilon\right) / \int \mathrm{d} \gamma_{\mathrm{e}} P\left(\gamma_{\mathrm{e}}, \epsilon\right)$ is the normalized pair distribution. The number of absorbed TeV photons $n_{\mathrm{TeV}}$ can be obtained by multiplying the intrinsic number distribution of VHE photons  by a factor of $1-{\rm e}^{-\tau_{\gamma \gamma}(\epsilon,z)}$, where $\tau_{\gamma \gamma}(\epsilon,z )$ is the optical depth due to the EBL absorption. Here we use the EBL model from \cite{saldana_2021}, in accordance with \cite{LHAASO_2023}. 


Due to the deflection effect of charged particles while travelling in the magnetic field, the arrival time of secondary photons from electromagnetic cascade will be delayed relative to the survival primary photons. Considering the scatter geometry, \cite{ichiki_2008} gave the time delay in the observer’s frame as 
\begin{equation}
\Delta t_{\mathrm{B}}\left(\gamma_{\mathrm{e}}, \epsilon,\mathrm{B} \right) \approx \frac{\left[\lambda_{\gamma \gamma}\left(\epsilon\right)+ \lambda_\mathrm{IC}\left(\gamma_\mathrm{e}\right)\right]\theta_\mathrm{B}^2(\gamma_\mathrm{e},\mathrm{B})}{2c} \label{eq:t_delay}
\end{equation}
where $\lambda_{\gamma \gamma}(\epsilon)$ is the mean free path of the TeV photons,  $\lambda_{\gamma \gamma}(\epsilon) = D /\tau_{\gamma \gamma}(\epsilon,z ) $ ($D$ is the distance of the GRB), and $\lambda_\mathrm{IC}(\gamma_\mathrm{e})$ is the cooling length of relativistic pairs with Lorentz factor $\gamma_\mathrm{e}$ due to inverse-Compton (IC) scatterings. Here $\theta_\mathrm{B}(\gamma_\mathrm{e},\mathrm{B})$ is the deflection angle, which depends on the coherence length $\lambda_\mathrm{coh}$, the strength $\mathrm{B}$ of the magnetic field and the energy of pairs $\gamma_\mathrm{e}$. Assuming that the magnetic field strength is constant and the coherence length $\lambda_\mathrm{coh}$ is larger than $\lambda_\mathrm{IC}(\gamma_\mathrm{e})$, the deflection angle $\theta_\mathrm{B}(\gamma_\mathrm{e},\mathrm{B})=\lambda_\mathrm{IC}(\gamma_\mathrm{e})/\mathrm{R}_\mathrm{L}(\gamma_\mathrm{e},\mathrm{B})$. If the coherence length $\lambda_\mathrm{coh}$ is smaller than $\lambda_\mathrm{IC}(\gamma_\mathrm{e})$, 
the behavior of electrons can be described as the random walk, so $\theta_\mathrm{B}(\gamma_\mathrm{e},\mathrm{B})=\sqrt{\lambda_\mathrm{coh}\lambda_\mathrm{IC}(\gamma_\mathrm{e})}/\mathrm{R}_\mathrm{L}(\gamma_\mathrm{e},\mathrm{B})$, where $\mathrm{R}_\mathrm{L}(\gamma_\mathrm{e},\mathrm{B})$ denotes the Larmor radius. 

Then the $e^+e^-$ pairs lose their energy through repeated scatterings on CMB photons with a cooling length of a few kpc ($e+\gamma_0\rightarrow{e+ \gamma_{2 \mathrm{nd}}}$).
The spectrum of the delayed emission can be obtained from the total time-integrated flux of $e^{\pm}$ responsible
for the delayed emission observed at time $t_{\mathrm{obs}}$ after the burst. 
Considering the effect of deflection in IGMF, previous works \citep{dai_2002-1,razzaque_2004,veres_2017,wang_2020} have employed a simple approximation to relate the number distribution of electrons used in calculating IC flux ($\mathrm{d} N/\mathrm{d} \gamma_{\mathrm{e}}$) with the  distribution of electrons generated by pair production ($\mathrm{d} N_{\mathrm{e},0}(\epsilon)/\mathrm{d} \gamma_{\mathrm{e}}$):
\begin{align}
\frac{\mathrm{d} N}{\mathrm{d} \gamma_{\mathrm{e}}} =t_{\mathrm{IC}}\left(\gamma_{\mathrm{e}}\right) \int \mathrm{d} \epsilon \frac{\mathrm{d} N_{\mathrm{e},0}(\epsilon)}{\mathrm{d} \gamma_{\mathrm{e}}} \frac{1}{\max \left(\Delta t_{\mathrm{B}}\left(\epsilon, \gamma_{\mathrm{e}}, B\right), \Delta t_{\mathrm{obs}}\right)}.\notag\\
\end{align}
This assumes that all photons that were emitted by electron pairs during the IC cooling time $t_{\mathrm{IC}}$ have been received by observer for a typical delay time $\Delta t_{\mathrm{B}}$ .

However, a more sophisticated treatment should consider the effect of cooling  on the the energy spectrum of pairs, which leads to a shift in population of electrons from high energy to low energy.
In this study, we take into account the evolution of electron distribution ($\partial N(\gamma_e, t) / \partial t=-\partial[\dot{\gamma_e} N(\gamma_e, t)] / \partial \gamma_e$). Consequently the distribution of electrons and time delay become time-dependent ($\mathrm{d} N_{\mathrm{e},0}(\epsilon)/\mathrm{d} \gamma_{\mathrm{e}}\to\mathrm{d} N_{\mathrm{e},\mathrm{t}}(\epsilon)/\mathrm{d} \gamma_{\mathrm{e}}$ ; $\Delta t_{\mathrm{B}}\left(\epsilon, \gamma_{\mathrm{e}},\mathrm{B} \right)\to\Delta t_{\mathrm{B}}\left(\epsilon, \gamma_{\mathrm{e}},\mathrm{B},\mathrm{t} \right)$). 
As a result, the electron distribution is modified to be 
\begin{align}
\frac{\mathrm{d} N}{\mathrm{d} \gamma_{\mathrm{e}}} = \int dt \int \mathrm{d} \epsilon \frac{\mathrm{d} N_{\mathrm{e},\mathrm{t}}(\epsilon)}{\mathrm{d} \gamma_{\mathrm{e}}} \frac{1}{\max \left(\Delta t_{\mathrm{B}}\left(\epsilon, \gamma_{\mathrm{e}},\mathrm{B},\mathrm{t} \right), \Delta t_{\mathrm{obs}}\right)}\notag\\
\end{align}
and the spectrum of IC photons is given by
\begin{equation}
\frac{d^{2} N_{\mathrm{delayed}}}{d E d t} =\int \frac{d \epsilon _0}{\epsilon _0} n(\epsilon _0)\int d \gamma_{e}\frac{3 \sigma_{T} c}{4 \gamma_{e}^{2}} f(x)\frac{\mathrm{d} N}{\mathrm{d} \gamma_{\mathrm{e}}}
\end{equation}
where $f(x)=2 x \ln x+x+1-2 x^{2}$ and $x=\epsilon_{\mathrm{IC}} / 4 \gamma_{e}^{2}\epsilon _0$. 
Additionally, it is worthy noting that the resulting photon flux represents an average effect over the entire observation period $\Delta t_{\mathrm{obs}}$, which spans from $ t_{\mathrm{start}}$ to $ t_{\mathrm{end}}$. If the time delay of the secondary photon emission is less than the start time of the observation period (i.e., $\Delta t_{\mathrm{B}}\left(\epsilon, \gamma_{\mathrm{e}}, \mathrm{B},\mathrm{t}\right) < t_{\mathrm{start}}$), then all corresponding secondary photons  have already arrived, so we remove the contribution from these particles in the calculation.

\subsection{Limits on IGMF using GRB 221009A} 
\begin{figure}
   \centering
   \includegraphics[angle=0,scale=0.35]{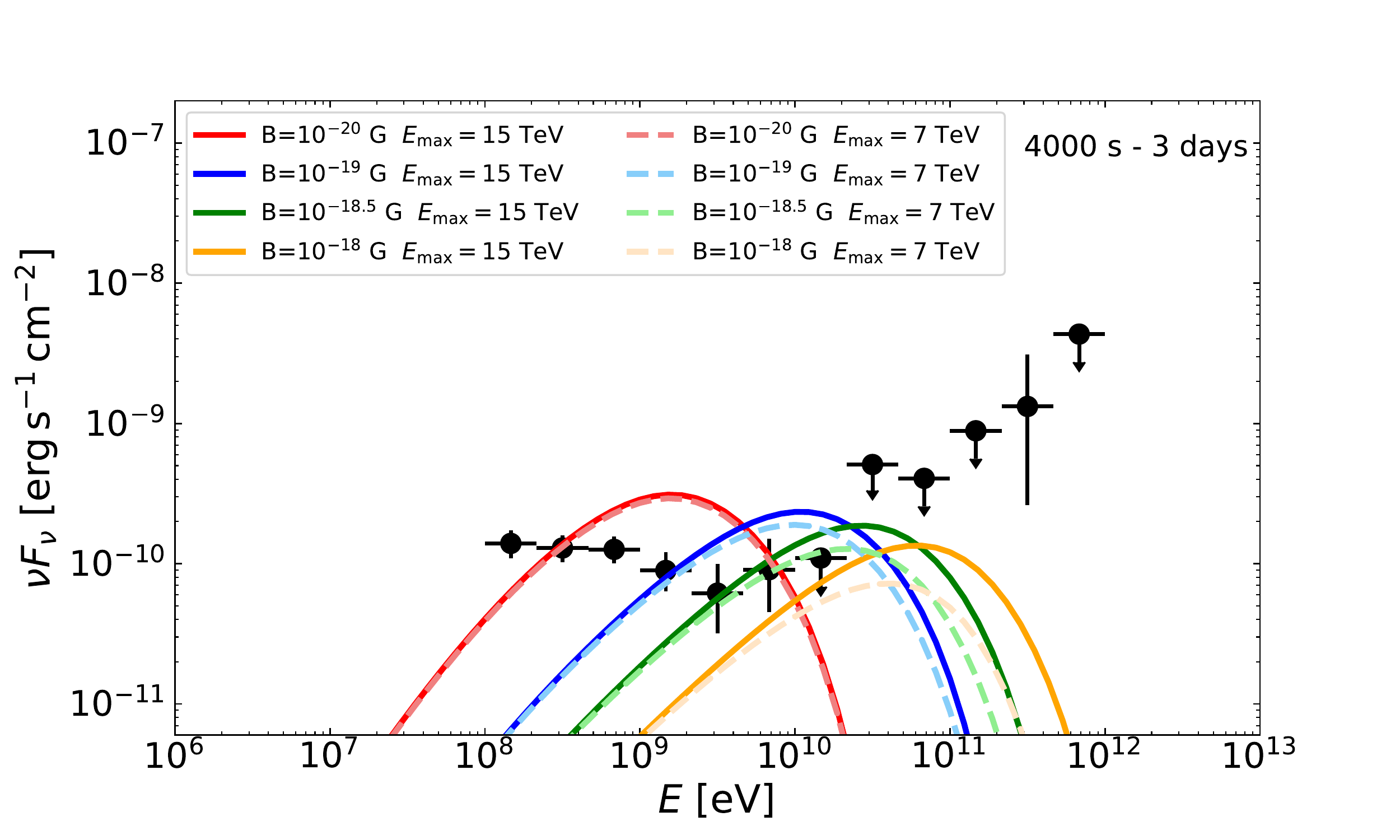}
   \includegraphics[angle=0,scale=0.35]{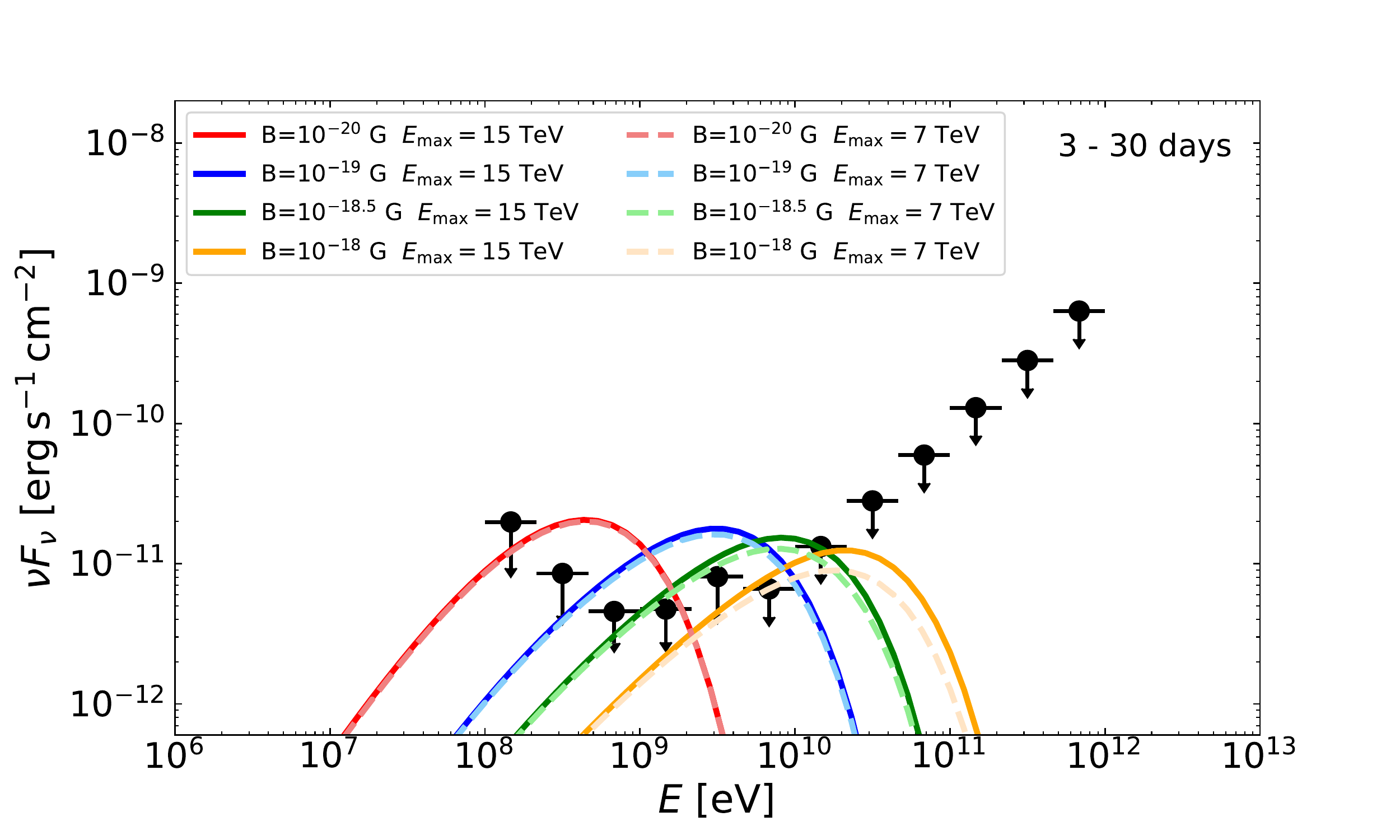}
   \includegraphics[angle=0,scale=0.35]{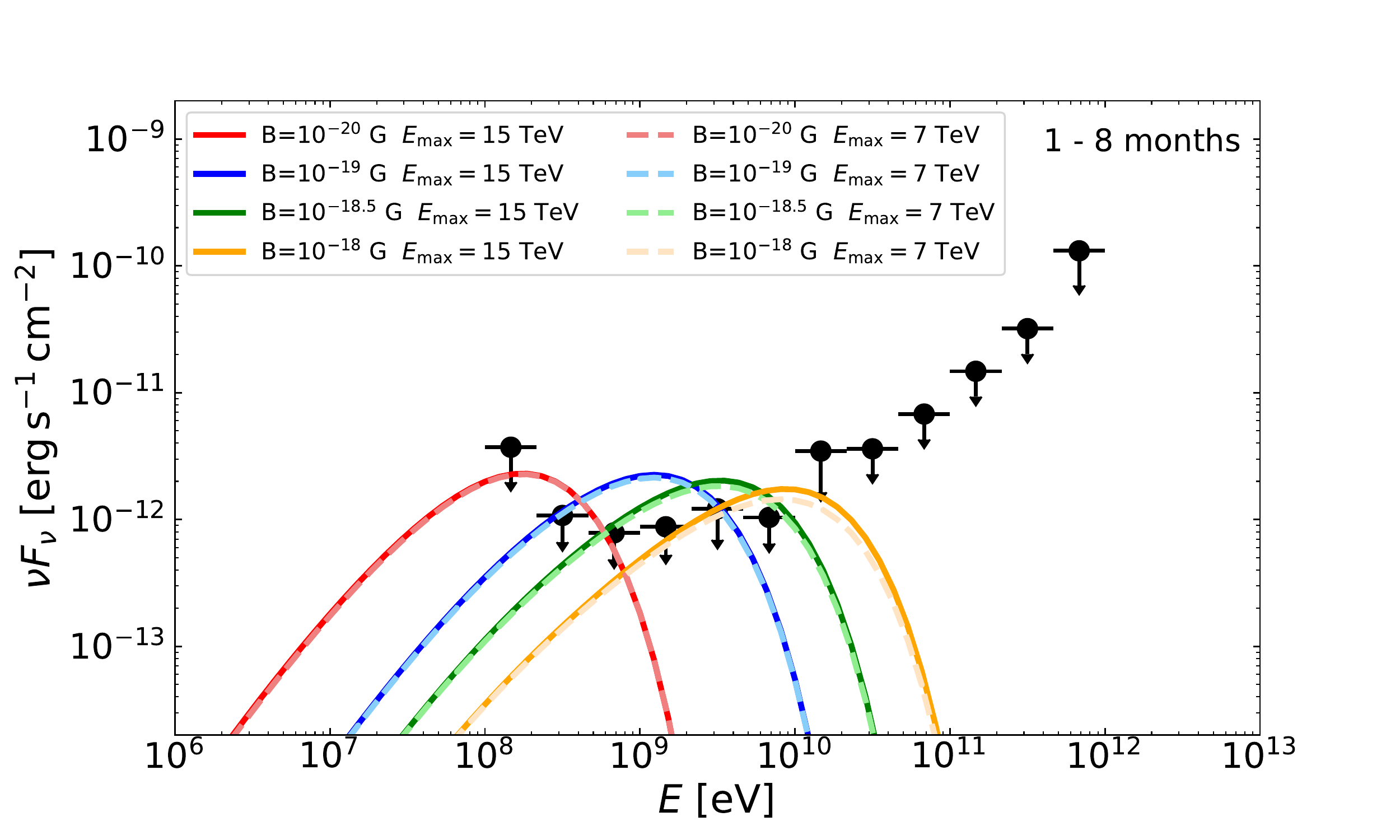}
   \caption{Comparison between the expected flux of the echo emission and the observed data by  Fermi/LAT in the analytic approach. The SEDs of the echo emission are averaged over the observation time 4000s-3 days (top panel), 3-30 days (middle panel) and 1-8 months (bottom panel). The dashed and solid lines represent the cases assuming a maximum energy of 7 TeV and 15 TeV, respectively. The black points denote the Fermi-LAT detection while the black arrows denote the upper limits.}
   \label{fig:analytic}
\end{figure}

Here, we consider four cases of the IGMF strength, ranging from $10^{-20}$ to $10^{-18}$ Guass. With the analytic description provided, we  have better understanding on the evolution trend of the cascade emission spectrum for varying magnetic field strengths as discussed in Section \ref{sec:Simulation}. According to Eq.\ref{eq:t_delay}, the arrival time is proportional to magnetic field strength and inversely proportional to the energy of primary photons. As a result, the peak energy of the cascade spectra varies systematically with the magnetic field strength and observing time, as shown in Fig.\ref{fig:analytic}. The favorable choice for studying the magnetic strengths is to ensure that the peak energy falls within the Fermi-LAT energy window. Specifically, choosing a relatively later time interval can provide a better constraint for the high IGMF cases.

In our calculation, the primary emission is treated as instantaneous injection. 
We assume that the shape of intrinsic spectrum is a power-law (i.e., $dN/dE = A(E/\mathrm{TeV})^{-\Gamma}$), which is consistent with the results reported in \cite{LHAASO_2023}. The normalization factor ($A= 1.56\times 10^{-7} ~\mathrm{TeV}^{-1} ~\mathrm{cm}^{-2} ~\mathrm{s}^{-1}$) is the average value of the time-resolved spectra over
the full TeV emission duration 1749 s, and the spectral index ($\Gamma=2.36$) is adopted from the value during the time interval when the TeV flux is dominant (from $\mathrm{T}_0+248$ s to $\mathrm{T}_0+326$ s). As there was no clear cutoff or break in the intrinsic spectrum below 7 TeV \citep{LHAASO_2023}, we set the maximum energy at 15 TeV, which we believe to be a reasonable choice. The coherence length of IGMF is set to be 1 Mpc. Figure \ref{fig:analytic} presents a comparison between the expected spectrum of the echo emission and the upper limit imposed by the Fermi/LAT observations. We can find that the predicted flux exceeds the upper limits imposed by Fermi/LAT for IGMF with $B\le 10^{-18}$ G, so we obtain a limit on IGMF of $B\ge 10^{-18}$ G for our analytic approach.

\end{document}